\documentclass[11pt]{article}

\usepackage{graphicx}
\usepackage{url}
\usepackage{hyperref}

\usepackage{cise-like}
\usepackage{enumitem}
\setdescription{font=\normalfont\sffamily\bfseries, itemsep=.5ex,
    parsep=.5ex, leftmargin=3ex}

\usepackage{xcolor}
\usepackage[normalem]{ulem}
\newcommand{\added}[1]{\textcolor{blue}{\uline{#1}}}
\newcommand{\deleted}[1]{\textcolor{red}{\sout{#1}}}
\renewcommand{\deleted}[1]{}
\renewcommand{\added}[1]{}

\begin{document}

\title{Mayavi: a package for 3D visualization of scientific data}

\author{Prabhu Ramachandran
\thanks{
Department of Aerospace Engineering,
IIT Bombay, Mumbai,
INDIA - 400 076
}
\and
Ga\"{e}l Varoquaux
\thanks{
Parietal team, INRIA Saclay Ile de France, Parc Orsay Universit\'{e}, 4 rue
J. Monod, 91893 Orsay FRANCE
}
\thanks{
NeuroSpin/CEA Saclay, B\^{a}t 145, 91191 Gif-sur-Yvette FRANCE
}
}

\date{}
\maketitle

\section{Introduction}

Mayavi is an open-source, general-purpose, 3D scientific visualization
package.  It seeks to provide easy and interactive tools for data
visualization that fit with the scientific user's workflow. 
\added{Mayavi is unique as it provides a consistent tool to address
visualization needs ranging from a turn-key application to building
blocks for domain-specific needs.} For this
purpose, Mayavi provides several entry points: a full-blown interactive
application; a Python library with both a MATLAB-like interface focused
on easy scripting and a feature-rich object hierarchy; widgets associated
with these objects for assembling in a domain-specific application, and
plugins that work with a general-purpose application-building framework.

\deleted{%
Mayavi draws much functionality from the integration of a tall stack of
powerful tools. While this may render the installation challenging,
various scientific Python distributions (\url{http://pythonxy.com},
\url{http://www.enthought.com/products/epd.php}) provide environments
for scientific computing with Python, and include Mayavi. In addition,
Mayavi is now included in the main Linux distributions. Finally, Mayavi
is distributed under the terms of the BSD license, in order to encourage
reuse by all kinds of users, including commercial products.
}

In this article, we present an overview of the various features of Mayavi, we
then provide insight on the design and engineering decisions made in
implementing Mayavi, and finally discuss a few novel applications.

\section{What is Mayavi?}

\subsection{A unique integration with a scientific workflow}

Mayavi is unlike most other visualization tools through a combination of
factors. First of all, the Mayavi project avoids domain-specific cases
and strives to build reusable, general-purpose, abstractions.  This is
important since different research fields often need to solve similar
problems. Second, Mayavi exposes tools and objects that fit closely with
a naive user's expectation, and also allows expert users to handle
complex visualizations. Finally, and most important, the project is more
than just a visualization library since it provides, in addition to the
library, widgets, dialogs, plugins, and an application, to fit in
various aspects of the scientific workflow.   A rich interactive
graphical application, along with simple but full-featured scripting is
important because visualization of complex datasets is best done
interactively, as the user can examine the data visually, tweak
parameters and build the visualization to suit the data. In addition, to
enable batch processing and non-interactive use, Mayavi visualizations
can be driven without a user interface.  The use of the Python language
is central to answering multiple scientific use cases, as it is a
powerful, yet easy-to-learn, programming language \cite{CiSE07}. Thanks
to a growing number of high-quality scientific libraries \cite{TO06,
JOP01}, Python has recently garnered significant mind-share among
scientists. Indeed, it lends itself to interactive use with simultaneous
plotting \cite{PG07, Hun07} that is ideal for scientific and
data-processing development. Mayavi brings powerful 3D scientific data
visualization to this tightly-integrated environment.

There are several excellent general-purpose 3D visualization programs
that expose high-quality Python interfaces, such as
ParaView~\cite{And06} or VisIt~\cite{Han05}. Unlike Mayavi, these tools
\added{transparently} support parallel data visualization as well as 4D
datasets.  However, Mayavi differs from these tools by a tighter
integration with the workflow of a typical scientist using Python for
numerical computation: it uses familiar data structures and exposes all
its graphical user interfaces as \deleted{reusable} components
\added{that can be used outside of the main application}. First of all,
Mayavi operates naturally on {\tt numpy} arrays, the core data structure
used in the main scientific Python projects. As detailed below, the
entire VTK API is wrapped with implicit conversion between VTK arrays
and {\tt numpy} arrays and the {\tt mlab} interface provides convenient
functions to visualize 3D data described by {\tt numpy} arrays. This,
along with the fact that Mayavi integrates well with IPython, makes it
highly convenient for interactive work. Second, Mayavi relies on a
reflexive object model in which each object can be used as a component
to create custom dialogs that embed a Mayavi visualization. \added{Mayavi
provides a readily usable, full-blown interactive environment for 3D
data visualization and is part of a larger suite of open source tools
called the Enthought Tool Suite} \deleted{Together with a larger suite
of tools} (ETS~\cite{ETS})\added{.}  \added{Mayavi along with ETS}
enables users to create rich, interactive scientific applications that
support 2D and 3D plotting requiring only the knowledge of Python and
object-oriented programming. Thus Mayavi fills a valuable need in the
scientific computing ecosystem.

\subsection{Powerful underlying technologies}

Mayavi is part of the \deleted{Enthought Tool Suite} ETS \cite{ETS} and
builds upon a powerful stack of existing libraries.  In this section we
provide a very brief overview of these.

\begin{description}
\item[VTK]: The Visualization ToolKit \cite{VTK} is one of the best,
actively-developed, general-purpose, open-source, visualization and graphics
libraries available.

\item[Numpy]: The workhorse of scientific computing with Python is the
{\tt numpy}
array structure. This multidimensional numerical array transforms Python into
a high-level array language, similar to MATLAB. It is used as a common data
container in most scientific libraries and projects relying on Python.

\item[Traits]: The Traits library 
(\url{http://code.enthought.com/projects/traits}), 
is the cornerstone of the ETS. Traits
extend Python object attributes and provide an elegant mechanism for attribute
initialization, validation, delegation, notification (efficient callbacks on
attribute modification), and visualization (through dialogs using wxPython or
Qt4). Henceforth, we refer to an object with traits as a ``traited
object''.

\item[TVTK]:  The Traited VTK library (TVTK) is {an automatically
    generated wrapper of VTK combined} \deleted{combines VTK} with
    Traits\added{.} \deleted{by} This is done by wrapping VTK-Python
    objects and providing a traits-enabled API with a more ``Pythonic''
    feel \added{for the VTK objects}.  All C++-style getters and setters
    are replaced by traits.  VTK has its own C++-based array structures,
    and the VTK-Python bindings require tedious manipulations to copy or
    reference the data from {\tt numpy} arrays to VTK arrays.  TVTK
    provides a seamless API where VTK arrays are converted dynamically
    from and to {\tt numpy} arrays, using a view of the same memory
    where possible.  In addition, because of the use of Traits, all TVTK
    objects provide a default dialog to edit their properties.  TVTK is
    thus a very powerful object hierarchy that maps with some simple
    rules to VTK and forms the foundation of Mayavi.

\item[TraitsUI/Envisage]: The user interface of the end-user application relies
entirely on Traits and optionally on Envisage.  Envisage is an
application-building framework, \`{a} la the Eclipse framework
\added{(\url{http://www.eclipse.org})}, which is used
for the application. At its core, Envisage is a system for defining,
registering, and using plugins.  Both Traits and Envisage can use wxPython or
PyQt as a backend.  Thus Mayavi can be used with both toolkits, although
currently the wxPython backend is more developed and thus more mature.

\end{description}

\subsection{Packaging and licensing}

\begin{sloppypar}
\added{%
Mayavi draws much functionality from the integration of a tall stack of
powerful tools. While this may render the installation challenging,
various scientific Python distributions (\url{http://pythonxy.com},
\url{http://www.enthought.com/products/epd.php}) provide environments
for scientific computing with Python, and include Mayavi. In addition,
Mayavi is now included in the main Linux distributions. Finally, Mayavi
is distributed under the terms of the BSD license, in order to encourage
reuse by all kinds of users, including commercial products.
}
\end{sloppypar}

\subsection{A simple pipeline model for visualization}

VTK relies on an elaborate pipeline model that is assembled to create a
visualization. In an effort to \deleted{simplify the model presented}
\added{present a simpler model} to the user, Mayavi exposes a basic
pipeline interface: data is obtained from a {\em data source}, and the
user can add visualization {\em modules} to display the data in various
ways, or {\em filters} to modify the data before eventual visualization.
Mayavi connects the various components of the pipeline and, to an
extent, also takes care of which components can be inter-connected.  For
example, a cloud of scattered points cannot be visualized with a grid
plane since there is no fixed grid for these points.  The different {\em
sources}, {\em modules} and {\em filters} are traited objects that all
have associated dialogs and methods that expose the underlying TVTK
objects as much as possible without requiring a detailed knowledge of
VTK.  The Mayavi pipeline collapses together several elements of a VTK
pipeline, for instance the Mayavi {\em modules} are made of VTK {\em
mappers}, {\em actors}, and eventually {\em widgets} or even {\em
filters} when their use helps representing the data. Finally, although
loops in the pipeline are possible, Mayavi \added{represents}
\deleted{exposes} the pipeline as a tree rather than a graph (see Figure
\ref{fig:mayavi}), to facilitate its manipulation, both through the
interactive user interface, and programmatically.

\subsection{History of the project}

Mayavi was created in 2001 \cite{Mayavi01} by P.~Ramachandran as an
end-user application for scientific visualization. The name is a
Sanskrit word meaning ``magician''. The application was appreciated for
its ease of use and interactivity.  However, it was not easy to script
from Python.  During 2004--2005, Enthought Inc.\  hired P.~Ramachandran
to write TVTK \cite{TVTK05} and start work on a new version of Mayavi.
The new version, ``Mayavi2'' \cite{Mayavi05}, uses the tools developed
at Enthought to focus on reuse and embedding.  In 2007, G. Varoquaux
joined the project.  The project is rapidly gaining features and
documentation while the usability is improved based on user feedback.
In early 2008, TVTK and Mayavi were each awarded FOSS India awards.

\section{Using Mayavi}

Mayavi can be used in different ways, and as such has several entry points.

\subsection{The {\tt mayavi2} application}

The interactive application, {\tt mayavi2}, is an end-user tool that can
be used without any programming knowledge. It provides an interface with
menus and several panels to guide the user while creating a
visualization (see Figure 1).  As described earlier in the overview,
Mayavi presents a simplified pipeline view of the visualization. Through
the menus, data can be loaded from files, or created with predefined
objects such as a layout of the Earth's continents, parametric surfaces
defining for example a Klein bottle, etc.  Subsequently, optional
filtering may be performed on the data, and visualization modules are
added to create the visualization.

\begin{figure}
\includegraphics[width=\linewidth]{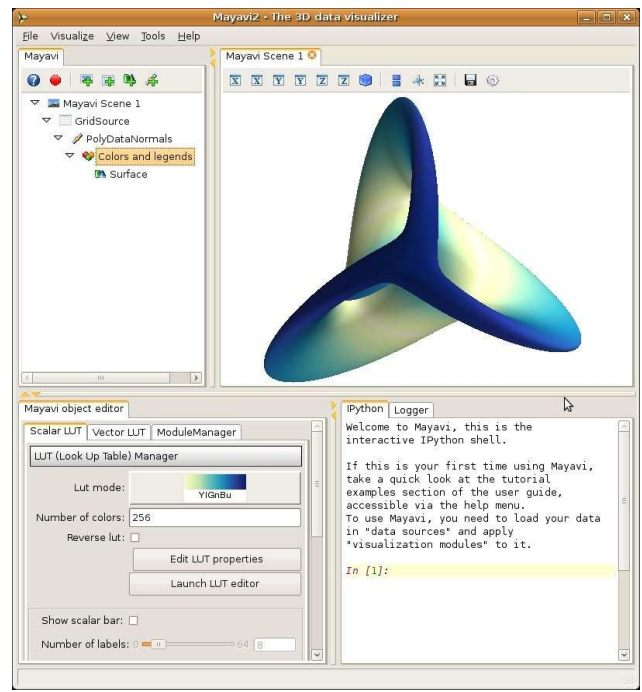}
\caption{
    A view of the {\tt mayavi2} application showing the Mayavi logo: a Boy
    mathematical surface. On the bottom right, an interactive IPython prompt,
    for prototyping scripts or exploring objects, can be seen. On the upper
    left stands the tree representing the visualization pipeline. It features
    a source, here a grid of points created by {\tt numpy} arrays, a
    {\tt PolyDataNormal}
    filter used to calculate normals of the surface for a smooth rendering,
    and the {\tt Surface} module, with its corresponding module-manager node, used
    to change the colors and legend of the surface. The dialog below the tree
    can be used to modify the selected node, i.e.\ colors and legend properties
    in the present case.
\label{fig:mayavi}
}
\end{figure}

The various pipeline components (sources, modules and filters) appear on
a {\em tree view} (see the left panel in Figure \ref{fig:mayavi}), and
more can be added through menus and dialogs. In particular, contextual
menus suggest to the user the filters or modules that are applicable to
a given data source.  The pipeline may also be reorganized using
drag-and-drop operations on the tree nodes.  Objects selected on the
pipeline view can be edited in another panel (left side of the bottom
panel in Figure \ref{fig:mayavi}) and modifications are immediately
applied to the visualization.  It is to be noted that although the
pipeline shown in the Figure \ref{fig:mayavi} is extremely simple, more
complex pipelines can also be setup.

While a raw VTK dataset is a versatile data structure describing data
embedded in a 3D space, a Mayavi source tries to expose to the user a
simple interface to importing data in Mayavi.  Similarly, the Mayavi
modules are a single point of entry to changing all the properties of an
object displayed on the visualization, and gather all the VTK
sub-objects in one object and one dialog.  An exception to this rule
is that the color maps and the legends can be shared between modules,
and thus can be represented as a separate node in the pipeline. 

\added{While TVTK is a complete wrapping of VTK,} Mayavi offers through
its primary tree-based interface only a limited subset of \added{all of}
VTK's filters, and the Mayavi sources cannot cover all possible ways to
create VTK datasets. This is why Mayavi offers a {\tt UserDefined}
filter to plug in almost any VTK filter in the Mayavi pipeline by
specifying its name, and a {\tt VTKDataSource} class to create a Mayavi
source from any VTK dataset. 

\begin{sloppypar}
The entire visualization can be saved to disk in a Mayavi-specific
format although, this is one aspect of Mayavi that needs to be more
robust.  Alternatively, the data generated at any point of the pipeline
can be saved to VTK files.  \added{We note that Mayavi supports the
native VTK datasets, which includes image data, structured grids,
polygonal datasets and unstructured grids.  Detailed information on the VTK
datasets is available from
\url{http://vtk.org/VTK/img/file-formats.pdf}}.
    
\end{sloppypar}

The {\tt mayavi2} application has a couple of features to help create Python
scripts from a visualization.  First, the application displays an interactive
Python shell, where Python commands can be entered for immediate execution.
The scripting API described later in this article can be used to create or
modify visualizations. Objects dragged from the pipeline tree to the shell
appear in it as Python objects for exploration or modification.  Second, the
pipeline tree view features a record button (the red button on the toolbar in
the left panel of Figure \ref{fig:mayavi}). When the record mode is turned on, any
modification to objects in the pipeline, or any object added to the pipeline
automatically generates the necessary lines of Python code to reproduce the
action.  Fully-functional Python scripts are thus created when the record mode
is turned on before adding any object to the pipeline, although the generated
code is not always the simplest possible code. In addition, the record mode is
also an extremely valuable learning tool for scripting Mayavi or TVTK objects.
Indeed, VTK is a very rich visualization library and Mayavi's object hierarchy
can be deep and complex. Thus, it can be hard to find the method or attribute
corresponding to a given feature. Consequently, the record feature is highly
convenient even for experienced VTK users.

An extensive user manual is shipped with Mayavi \cite{Manual} and is accessible
from the application. The user guide is rendered using Sphinx
(\url{http://sphinx.pocoo.org/}) and embeds a search bar and an index.

\subsection{Simple Python scripting}

As we have seen, Mayavi can be used in a fully-interactive manner, by a
non-programmer, using the {\tt mayavi2} application.  However, Mayavi can also be
used through a simple and yet powerful scripting API, providing a workflow
similar to that of MATLAB or Mathematica. Many scientists use Python for their
computational work.  Visualization is a very important component of such work
and is most effective when used interactively, as an exploratory and debugging
tool.  A good example of this is the typical use of matplotlib
\cite{Hun07} and IPython \cite{PG07}.

\begin{figure}
\includegraphics[width=\linewidth]{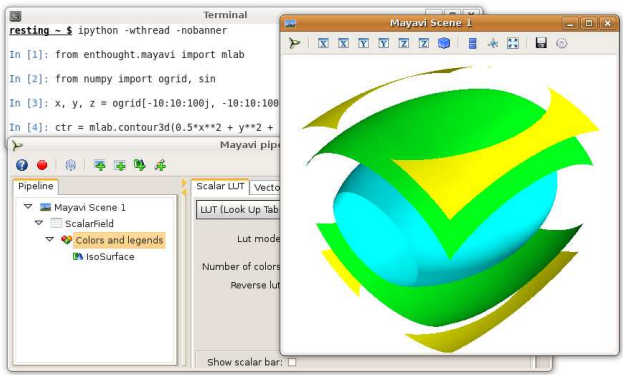}
\caption{
    Working with Mayavi in IPython. The terminal on the background runs the
    IPython session from which the visualization window on the foreground
    (right) was created. The pipeline dialog editing the different
    visualization objects was created by clicking on the button with the
    Mayavi icon on the left of the visualization window's toolbar.
\label{fig:ipython}
}
\end{figure}

Mayavi's {\tt mlab} scripting interface is a set of Python functions
that work with {\tt numpy} arrays and draw some inspiration from the MATLAB and
matplotlib plotting functions. It can be used
interactively in IPython, or inside any Python script or application.  The
following example generates iso-contours of a mathematical function, sampled
on a regular grid. The resulting visualization can be seen in Figure
\ref{fig:ipython}.

\begin{verbatim}
from enthought.mayavi import mlab
from numpy import ogrid

x, y, z = ogrid[-10:10:100j, -10:10:100j, -10:10:100j]

ctr = mlab.contour3d(0.5*x**2 + y**2 + 2*z**2)
mlab.show()
\end{verbatim}

Simple plotting commands operating on {\tt numpy} arrays, such as {\tt 
mlab.contour3d},
used in the previous example, build a complete visualization pipeline. These
simple commands hide the pipeline model from the user, for simple use cases.
They accept a large number of extra arguments to control the properties of the
visualization created. In addition, as they return the Mayavi modules created,
more properties can be changed by modifying their attributes.

Alternatively, {\tt mlab} offers a direct control of the Mayavi pipeline through
separate creation of sources, filters and modules. Thus the call to
{\tt mlab.contour3d} in the previous example can be replaced by two commands, one
to create a source object from the regularly-spaced volumetric data in a
{\tt numpy} array, and a second to apply an iso-surface module on it:

\begin{verbatim}
src = mlab.pipeline.scalar_field(0.5*x**2 + y**2 + 2*z**2)
ctr = mlab.pipeline.iso_surface(src)
\end{verbatim}

Manually populating the pipeline requires an understanding of the pipeline
model.  It is also more powerful as it gives access to a wider range of
possibilities through custom-made pipelines. The names of the {\tt 
mlab.pipeline}
functions to create objects are lower-case-with-underscores versions of the
camel-case names of the classes represented, as they appear by default in the
pipeline view. Thus, going from a pipeline built interactively to a script is
very easy.

Although the window used to display visualization is very simple, the full
power of Mayavi is still available.  Clicking the button with the Mayavi logo
on the left of the visualization-window toolbar displays a dialog containing
the same pipeline tree view as in the Mayavi application (see Figure
\ref{fig:mayavi}). All
the interactive functionality of the {\tt mayavi2} application is accessible in
this dialog: the buttons on the toolbar provide access to help, script
recording, or object-creation. Clicking on the pipeline nodes creates dialogs
that allow modification of the object properties. The pipeline can be
populated and modified by the context menus accessible with a right-click on
the nodes.

\subsection{Animating data and building interactive dialogs}

\begin{figure}
\includegraphics[width=\linewidth]{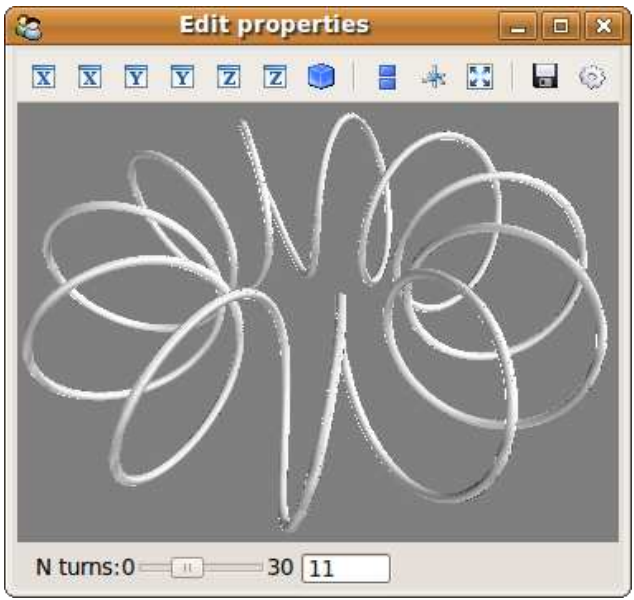}
\caption{
    The dialog created by the code example.  In addition to the interaction
    with the visualization, dragging the slider bar modifies the
    visualization interactively.
\label{fig:interactive}
}
\end{figure}

Objects created by the {\tt mlab} functions expose an {\tt mlab\_source} attribute,
which gives access to the {\tt numpy} arrays used to create the dataset.
Assigning new arrays to the {\tt mlab\_source} triggers an update of the
visualization. The name of the attribute one needs to modify on the
{\tt mlab\_source} object is the name given to the corresponding argument in the
function signature documentation, for instance the {\tt mlab.contour3d} function
signature is {\tt contour3d(scalars, ...)}, where {\tt scalars} is a 3D array. Thus,
we can animate the contour object created previously by modifying in place the
scalars it represents:

\begin{verbatim}
from time import sleep
for i in range(1, 10):
    sleep(1)
    ctr.mlab_source.scalars = 0.5*x**2 + y**2 + i*z**2
\end{verbatim}

In-place modifications are also useful when embedding a visualization in an
interactive application. A Mayavi scene can be displayed as part of a
traits-based user interface. The following example displays a dialog
visualizing a 1D parametric function embedded in a 3-Dimensional space as a
curved line. The mathematical curve, defined by the {\tt curve} function, takes
one parameter -- the number of minor rotations in the transverse direction.
The dialog (Figure \ref{fig:interactive}) enables the modification of this parameter with an
immediate visualization of the results:

\begin{verbatim}
from numpy import linspace, pi, cos, sin
from enthought.traits.api import HasTraits, Range, Instance, \
    on_trait_change
from enthought.traits.ui.api import View, Item, HGroup
from enthought.mayavi.core.ui.api import SceneEditor, \
    MlabSceneModel

def curve(n_turns):
    phi = linspace(0, 2*pi, 2000)
    return [ cos(phi) * (1 + 0.5*cos(n_turns*phi)),
    	sin(phi) * (1 + 0.5*cos(n_turns*phi)),
    	0.5*sin(n_turns*phi)]

class Visualization(HasTraits):
    n_turns = Range(0, 30, 11)
    scene   = Instance(MlabSceneModel, ())

    def __init__(self):
        HasTraits.__init__(self)
        x, y, z = curve(self.n_turns)
        self.plot = self.scene.mlab.plot3d(x, y, z)

    @on_trait_change('n_turns')
    def update_plot(self):
        x, y, z = curve(self.n_turns)
        self.plot.mlab_source.set(x=x, y=y, z=z)

    view = View(Item('scene', height=300, show_label=False,
    		editor=SceneEditor()), 
    	    HGroup('n_turns'), resizable=True)

Visualization().configure_traits()
\end{verbatim}

In the above code example, the {\tt Visualization} class defines a few traits
including the {\tt scene} trait which is an instance of {\tt
MlabSceneModel}. The
{\tt configure\_traits()} call at the end of the code creates a dialog representing
the object, the layout of which is given by the {\tt view} defined at the end of
the class.  This view exposes the {\tt scene} trait using a {\tt SceneEditor} in the
dialog, and the {\tt n\_turns} attribute as a slider. On creation of the
{\tt Visualization} object, the curve is plotted in the embedded scene with the
{\tt plot3d} mlab call. When the {\tt n\_turns} attribute is modified, the 
{\tt update\_plot}
method is called, curve data is recomputed, and the plot object is modified
using the {\tt mlab\_source} attribute.

In general, the different properties of the objects used in visualizations,
such as sources, filters, modules, or even the scenes, can be modified in a
script, with instantaneous or delayed update of the scene, by simply setting
the corresponding trait. Moreover, the different dialogs that form Mayavi,
such as the tree view, or the dialog editing an object, can be embedded in a
user's application. Callbacks between these different dialogs and the scene
are already wired. As such, Mayavi forms more than a visualization library; it
can be used as a set of interactive components to provide live visualization
to a domain-specific application, requiring little knowledge of GUI
programming or event-loops.

\subsection{Embedding in existing applications}

Although Traits is a very powerful tool for developing interactive
applications, most existing applications are developed using a raw GUI
toolkit. It is thus important to integrate the dialogs produced from the code
in the previous paragraphs in a non Traits-aware GUI.  Traits has a wxPython
and a Qt4 backend. While the {\tt configure\_traits} method used in the above
example to create the dialog creates a full wxPython application and starts
the main event loop, the {\tt HasTraits} class also provides an {\tt 
edit\_traits}
method that only creates and returns a panel or dialog. Below is an example
showing how the {\tt Visualization} class defined earlier, and the corresponding
dialog, can be embedded in a wxPython application:

\begin{verbatim}
import wx

class MainWindow(wx.Frame):
    def __init__(self, parent, id):
        wx.Frame.__init__(self, parent, id, 'Mayavi in Wx')
        self.visualization = Visualization()
        self.control = self.visualization.edit_traits(
                        parent=self, kind='subpanel').control
        self.Show()

app = wx.PySimpleApp()
frame = MainWindow(None, wx.ID_ANY)
app.MainLoop()
\end{verbatim}

In the above example, the {\tt edit\_traits} method is passed the wxPython frame
into which the dialog is embedded.  The {\tt control} trait of the object produced
by the {\tt edit\_traits} call is the wxPython object containing the widget.
Similarly, dialogs can be embedded in a PyQt application, as detailed in the
user guide.

Any Mayavi dialog can be embedded similarly in more complex applications. For
instance the various Mayavi pipeline objects also provide an {\tt 
edit\_traits}
method to edit their properties. Thus, the work invested in developing
powerful widgets for the Mayavi application, such as the pipeline tree view,
is readily available to the application builder. 

\subsection{Extending the {\tt mayavi2} application}
\label{sec:extending-mayavi}

Instead of creating a new application, one can extend the already-powerful
{\tt mayavi2} application by adding to it custom functionality or domain-specific
elements. The application is built by integrating the 3D visualization
provided by Mayavi with other functionality, such as a Python shell, via
Envisage plugins. Using the same mechanism, one may put together other
applications or extend the {\tt mayavi2} application. A discussion of the Envisage
application-building framework is beyond the scope of this article.

In addition, Mayavi has a mechanism to register new types of data sources,
filters or modules. These are automatically added to the various menus and the
{\tt mlab.pipeline} interface.  A common use case is to add domain-specific file
readers or visualization modules. For example, the following code can be placed
in a module imported in {\tt \~ \//.mayavi2/user\_mayavi.py} to define a reader using
{\tt numpy} to load arrays stored in a text file:

\begin{verbatim}
from enthought.mayavi.core.api import registry, \
                        SourceMetadata, PipelineInfo
from enthought.mayavi.sources.api import ArraySource
import numpy as np

def array_reader(fname, engine):
    return ArraySource(scalar_data=np.loadtxt(fname))

registry.sources.append(SourceMetadata(
    factory     = __name__ + '.array_reader',
    menu_name   = "Array text files",
    extensions  = ['txt'],
    wildcards   = 'TXT files (*.txt)|*.txt',
    output_info = PipelineInfo(datasets=['image_data'],
                            attribute_types=['any'],
                            attributes=['any']),
    ))
\end{verbatim}

Of course, proper use of Mayavi as a platform for domain-specific
applications requires a good understanding of the finer details of Mayavi,
which is beyond the scope of this article, but is detailed in the user
manual \cite{Manual}.

\section{Mayavi architecture and software design}

In this section we provide a broad overview of the architecture and
software design of Mayavi.

\subsection{Architecture overview}

\begin{figure}
\includegraphics[width=\linewidth]{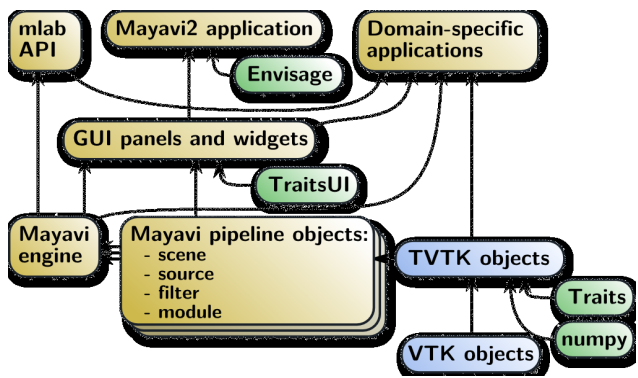}
\caption{
    Architecture diagram of Mayavi. The green nodes are external Python
    modules, the blue nodes are the VTK-based object hierarchy, and the
    yellow ones form the Mayavi functionality.
\label{fig:architecture}
}
\end{figure}

The general software architecture of Mayavi is summarized in the diagram on
Figure \ref{fig:architecture}. The visualization layer of Mayavi relies on TVTK objects.  The
Mayavi pipeline objects use TVTK objects and have methods that help wire them
together and simplify building the VTK pipeline. A central object, the Mayavi
{\tt Engine}, manages all the pipeline objects making up the visualization.

The set of pipeline objects and the engine form the core of the Mayavi
functionality. The mlab scripting API controls the engine to create
visualizations. As all objects rely on the Traits library, UI panels or
widgets can be created  using the TraitsUI package. The different panels, as
well as the core functionality of Mayavi, can be combined with other Envisage
plugins to create applications such as the {\tt mayavi2} application.

\subsection{The Engine as a pipeline warden}

The Mayavi engine maintains a tree structure of pipeline objects. Each
pipeline object maintains references to its parents and children, and exposes
a list of inputs and outputs.  The output list contains the TVTK datasets
flowing between the pipeline elements. The input objects are Mayavi pipeline
objects.  The Mayavi pipeline objects have callbacks to rewire the underlying
VTK pipeline if their inputs change.  They also feature two events,
{\tt pipeline\_changed} and {\tt data\_changed}, which are propagated down the pipeline
to update it. The engine manages the life-cycle of the pipeline objects, i.e.
it manages their addition and removal.

The pipeline of the iso-surface example introduced with the {\tt mlab} API in the
beginning of the article can be explicitly built with the following code: 

\begin{verbatim}
from enthought.mayavi.core.api import Engine
from enthought.mayavi.sources.api import ArraySource
from enthought.mayavi.modules.api import IsoSurface
from enthought.pyface.api import GUI
from numpy import ogrid
x, y, z = ogrid[-10:10:100j, -10:10:100j, -10:10:100j]

engine = Engine()
engine.start()
engine.new_scene()

src = ArraySource(scalar_data=(0.5*x**2 + y**2 + 2*z**2))
engine.add_source(src)
engine.add_module(IsoSurface())
GUI().start_event_loop()
\end{verbatim}

In the above, the {\tt engine} manages the connection between the source and
modules internally: it maintains the context of the visualization. For instance
the {\tt new\_scene} method can be overridden in a subclass to create an embedded
scene, as done in the {\tt mayavi2} application, or a separate window, as when
using {\tt mlab} in IPython. Other subclasses for off-screen rendering are also
available. In addition, the engine provides context-dependent actions which
can be useful to drive an interactive application in a manner similar to
spreadsheet scripting. 
\deleted{For instance, consider the case where there are two
sources. The {\tt current\_object} trait of the engine defines the source to which
subsequent filters and modules may be added.} 
\added{For instance, if there are several
sources, the {\tt current\_object} trait of the engine defines the source to which
 modules will be added if using the {\tt add\_module} method of the
engine.}
In the above, 
the {\tt
add\_module}
call implicitly adds the {\tt IsoSurface} instance to the \added{current
object: the} {\tt
ArraySource} source \deleted{,}
{\tt src}.  It is also possible to build the above pipeline explicitly by
selecting and connecting the different objects rather than delegating the task
to the engine:

\begin{verbatim}
scene = engine.current_scene
src = ArraySource(scalar_data=(0.5*x**2 + y**2 + 2*z**2))
scene.add_child(src)
src.add_child(IsoSurface())
\end{verbatim}

We note that the Engine is not global.  While the {\tt mayavi2}
application and {\tt mlab} provide default engines which suffice for
most of the use cases, a user can create many different engines for
different needs.  As a result, the scope, or context, of the different
visualizations and actions can be limited and controlled. This isolation
is important when one wishes to avoid side-effects in a large
application. It also makes Mayavi much easier to reuse and test.  For
example, if the Engine was global then any changes to it in a test suite
would influence other tests.  However, since one may create as many
engines as one desires it is easy to write tests that avoid unnecessary
side-effects.

\subsection{A central registry to avoid duplication}

Mayavi provides a large, and growing, list of pipeline objects. They are
exposed through many different interfaces, both graphical, and
programming APIs. In order to avoid code duplication, all the different
sources, filters, and modules are specified in a central registry along
with information describing their functionality.  The metadata
information in the registry is used to generate the different menus of
the user interfaces, as well as many of the {\tt mlab.pipeline}
functions, thus enforcing consistency throughout Mayavi. A sample of
this information can be seen in the example provided in
section~\ref{sec:extending-mayavi} where Mayavi is extended with a new
source by adding it to the registry.

\subsection{Model-view separation}

As much as possible, Mayavi uses a reactive programming style employing
callbacks on trait modifications.  UIs are created using representations
of the objects' traits. This programming style allows for a very clean
separation of the model from the view.  The model can be fully described
by the traits of all the objects on the pipeline.  For example, the
script-recording functionality described earlier is implemented in large
part by tracking the modifications to these traits.  Mayavi's use of the
Model View Controller (MVC) design pattern~\cite{Gamma} is not complete
and there is some mixing of the view and model.  However, while this can
be reduced in the future, we believe that the design is already very
reusable, offering us most of the advantages of MVC.

The ``message-passing'' style that replaces method calls by trait assignments is
thread-friendly and can be used to avoid dealing with GUI event loops. The
code updating the UI for instance is not exposed to the user.  As we mostly
rely on Traits for views, these are defined in a declarative way (as in the
interactive dialog example above).  Consequently, no application logic can be
found in the view-related code. This is very important for a clean scripting
API and also enables scripts to run in a headless (off-screen) mode with no
modifications \deleted{if using a suitable VTK build}
\added{when VTK is suitably built}.

\subsection{Testing and scriptable APIs}

As seen above, Mayavi strives to be highly reusable in a variety of
contexts for the user and developer.  One of the development strategies
that really helped make the API reusable and the code reliable is our
focus on being able to script the API as much as possible. This was
achieved in part by resorting to unit testing, integration testing, and
some Test-Driven Development (TDD) \cite{Beck}.  Testing and good
example scripts forced us to make the API highly reusable and resulted
in increased reliability and clean code. This experience corroborates
with the advantages claimed by practitioners of TDD.

\subsection{Summary of key design choices}

We believe some architectural decisions are key to the success of Mayavi's
reusability:

\begin{description}

\item[Multiple abstraction layers, as summarized by Figure 4.] This layered
  functionality enforces separation of concerns and enables sharing code
  between different entry points or APIs to address various use cases.

\item[A central, well-defined object: the Engine.] It coordinates the
    visualization, provides a context, and is thus important in
    establishing a consistent view of the application, both in the
    Mayavi code, and in the scripting APIs. In addition, this central
    object helps avoiding globals.  We note that any number of Engines
    can be created and used simultaneously, thus the Engine does not
    preclude the possibility of data-parallel execution.

\item[The use of Traits.] Traits' powerful object model leads to a good design 
  through reactive programming and strong model-view separation, in addition 
  to providing multiple-backend UIs with little effort.

\item[Model/view separation and loose coupling.] It is well known that 
  GUI-related
  code should be well separated from the core application logic.  However, in
  addition to this, we find that all helper-code unrelated to the core
  functionality which caters to common end-user needs, such as provided by the
  {\tt mlab} API, should also be separate from the core.
 
\item[Automatically-created objects and functions.] The TVTK wrapper code is
  entirely automatically generated.  Large parts of the UI-related boilerplate
  code, and some of the APIs, are auto-generated. This reduces duplication and 
  thus makes the interfaces more consistent, and the code easier to maintain.

\item[Focus on the API.] Striving for a simple API greatly improves the developers
  mental representation of the library and application's model, and as a
  result its architecture. The API should answer common use-cases and be
  consistent across the various needs.  Interaction and feedback from users,
  whose needs sometimes differ from that of the developers, has proven
  priceless.  
  
\item[Testing and scripting.] Unit tests, TDD and example scripts for users are
  invaluable in creating a truly reusable tool.  These practices are also a
  great way to notice unwanted tight coupling in the object model.

\end{description}

\section{Some real-word applications}

\subsection{Weather visualization using Mayavi}

\begin{figure}
\includegraphics[width=\linewidth]{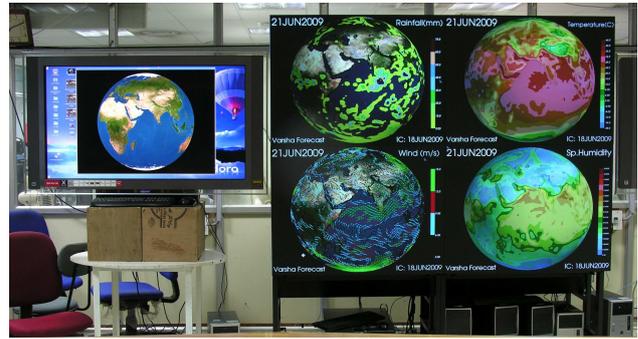}

\caption{
    Four Mayavi applications controlled via the network from a central program
    providing a synchronized view.
\label{fig:NAL}
}
\end{figure}

The FloSolver division at NAL (National Aerospace Laboratories,
Bangalore) uses Mayavi to visualize data produced by their weather
modeling code which is used to primarily forecast the monsoon in India.
Mayavi is used both as a display device and more importantly to help
refine the weather models. Data generated from weather simulations is
rendered interactively using Mayavi scripted from Python to automate
many mundane tasks.  The automatic scripting mode of Mayavi is used to
record UI actions and generates human readable Python code.  This code
is then hand-edited to produce visualizations of the weather data.
Thus, without a direct knowledge of Mayavi's internals or even a good
knowledge of Python, the scientists at NAL are able to generate
fully-working Python scripts and tailor them to their needs. For the
interactive display of the different atmospheric fields, each field is
displayed on a separate computer running a Mayavi script.  The camera
position of each Mayavi application is controlled by an in-house OpenGL
application used for display of cloud data obtained from satellite
images.  In order to do this, we wrote a simple TCP and UDP server
program that lets a user send Python statements across the network which
are interpreted by the running Mayavi application.  We used the
excellent Twisted (\url{http://twistedmatrix.com/trac/}) library for
this.  The size of the resulting server module was about 90 lines of
code (without the documentation/comments).  Using the server required
two additional lines of code in the existing scripts at NAL.  The
hardware setup at NAL is shown in Figure \ref{fig:NAL}.  The left-most
screen shows the application controlling the view of the 4 other Mayavi
applications via the network.  The right side is a visualization wall
consisting of 4 separate LCD panels put together as one.  This would not
be possible but for the powerful libraries available with Python.

\subsection{Web-browser based usage of Mayavi}

In a recent development, O. Certik and P. Ramachandran worked together
to setup a Sage (\url{http://www.sagemath.org}) notebook working with Mayavi in order to perform
visualizations on the web.  Sage provides, among many other things, a powerful
environment to do mathematics on the web through a notebook interface.  This
interface essentially provides a powerful Python-capable web page where users
can interact with a Python interpreter, and embed the results along with
images and text seamlessly from a browser window.

In order to have Mayavi working in the Sage notebook, we built VTK with
support for using Mesa's (\url{http://www.mesa3d.org}) OSMesa library for pure off-screen rendering.
Then by using the existing support in Mayavi for off-screen rendering we are
able to render images and generate X3D files, displayed interactively by the
Sage notebook. The web page \url{http://lab.femhub.org/home/pub/39} demonstrates the
resulting Sage notebook.  The page uses Mayavi's {\tt mlab} interface to generate
a visualization and produce an X3D file that may be visualized interactively
using a browser plugin.

\section{Conclusions}

There is a growing trend of moving more and more computational code to
high-level environments. Python is an increasingly popular
high-level language for scientific computing because it has the potential to
unite a variety of modules into a homogeneous environment.  

\added{In this article we have shown how} Mayavi provides a rich and powerful 3D
data visualization package that tightly integrates the various aspects
of scientific computing and scientific-application development in
Python. It strongly focuses on being reusable. In particular, it is
well-suited to applied science and engineering problems for which
building custom visualization tools is an important challenge. It tries
to match the different scientific computing work flows: interactively
with an end-user application, in scripts, or in custom applications and
also for pure off-screen rendering. The visualization model is
consistent throughout the various entry points and the interactive
application can be used to prototype visualizations that can be easily
converted into code embedded in scripts or applications.  Mayavi
dovetails nicely into a rich set of scientific tools using Python as a
natural, easy to use, 3D visualization environment.


\begin{thebibliography}{1}

\bibitem{CiSE07} {\em Computing in Science and Engineering} special
    issue on scientific computing with Python, vol. 09, no. 3, May/June,
    2007.

\bibitem{TO06} T. E. Oliphant, ``A Guide to NumPy'', Trelgol publishing, 
   \url{http://numpy.scipy.org}, 2006.

\bibitem{JOP01} E. Jones, T. E. Oliphant, P. Peterson, ``SciPy: Open source 
    scientific tools for Python'', \url{http://www.scipy.org}, 2001.

\bibitem{PG07} F. Perez, B. E. Granger, ``IPython: A System for Interactive 
   Scientific Computing'', {\em Computing in Science \& Engineering}, 2007.

\bibitem{Hun07} J. D. Hunter, ``Matplotlib: A 2D Graphics Environment'',
    {\em Computing in Science \& Engineering}, 2007.

\bibitem{And06} Andy Cedilnik, Berk Geveci, Kenneth Moreland, James
    Ahrens, and Jean Favre, ``Remote Large Data Visualization in the
    ParaView Framework'', In {\em Eurographics Parallel Graphics and
    Visualization} pp 163--170, May 2006.

\bibitem{Han05} Hank Childs, Eric Brugger, Kathleen Bonnell, Jeremy
    Meredith, Mark Miller, Brad Whitlock, Nelson Max, ``A Contract Based
    System For Large Data Visualization'', {\em 16th IEEE Visualization
    2005}, pp. 25, 2005.

\bibitem{ETS} Enthought, Inc, Austin, ``Enthought Tool Suite'',
   \url{http://www.enthought.com/products/ets.php}

\bibitem{VTK} W. \added{Schroeder} \deleted{Schrodeder}, K. Martin, 
    W. Lorensen, 
    ``The Visualization Toolkit'', Kitware, 4th edition, 2006.

\bibitem{Mayavi01} P. Ramachandran, ``MayaVi: A free tool for CFD data
   visualization.'' In 4th Annual CFD Symposium, Bangalore, India.  
   Aeronautical Society of India, August 2001.

\bibitem{TVTK05} P. Ramachandran, ``TVTK A Pythonic VTK'', EuroPython 
   Conference Proceedings, Goteborg, Sweden June 2005.

\bibitem{Mayavi05} P. Ramachandran. ``MayaVi2: The next generation'', 
   EuroPython Conference Proceedings, Goteborg, Sweden June 2005.

\bibitem{Manual} P. Ramachandran, G. Varoquaux, ``Mayavi Users Guide'',
   \url{http://code.enthought.com/projects/mayavi}, 2008.

\bibitem{Gamma} E. Gamma {\sl et al}, ``Design Patterns - Elements of 
    Reusable Object-Oriented Software'', Addison Wesley, 1994.

\bibitem{Beck} K. Beck, ``Test-Driven Development by Example'', Addison
    Wesley, 2003.

\end{thebibliography}
\end{document}